\documentclass[12pt, a4paper]{article} 

\usepackage[utf8]{inputenc}
\usepackage{amsmath, amssymb}
\usepackage{graphicx}
\usepackage{hyperref}
\usepackage{authblk} 
\usepackage{natbib}
\usepackage[left=2.5cm, right=2.5cm, top=4cm, bottom=4cm]{geometry}
\title{Maneuvering of an underwater vehicle using bio-inspired pectoral fins}

\author[1,*]{Pedro C. Ormonde}
\author[1,$\dagger$]{Xiaowei He}
\author[1]{Kenneth Breuer}

\affil[1]{School of Engineering, Brown University, Providence, RI 02912, USA}
\affil[*]{Correspondence: pco@brown.edu}
\affil[$\dagger$]{Current address: Department of Mechanical Engineering, University of Utah, Salt Lake City, UT 84112, USA}

\date{} 

\begin{document}

\maketitle

\begin{abstract}
A Cyber-physical underwater vehicle is equipped with bio-inspired flapping fins positioned on the sides of the vehicle's main body. The proposed control surfaces are inspired by fish pectoral fins, generating forces and moments that can potentially be harnessed for maneuvering, hovering and station keeping. The streamwise and cross-stream forces produced by the fins are characterized for a range of reduced frequencies and Strouhal numbers. The streamwise forces are shown to be predominantly a function of the fin's projected frontal area, while the lateral forces also depend on the Strouhal number.  When operated simultaneously, different flapping synchronizations can be employed for specific goals; a symmetric motion suppresses the lateral forces, while an anti-symmetric motion decreases the peaks of the streamwise force produced. The Cyber-physical vehicle demonstrates how the pair of fins can successfully maneuver the vehicle in the lateral direction.
\end{abstract}

\vspace{1em}
\noindent \textbf{Keywords:} Underwater locomotion; Bio-inspired maneuvering; Bio-locomotion; Pectoral fins

\section{Introduction}

Fish are highly maneuverable while also being energetically efficient swimmers ~\cite{LauderDrucker2004}. The majority of species employ body and caudal fin undulations as their primary source of thrust during high speed forward swimming~\citep{lauder2000function,webb1994biology}. While accessory appendages such as the pectoral fin may also be used for steady rectilinear propulsion~\citep{drucker1996pectoral,lauder1996pectoral,webb1973kinematics,westneat1996functional}, they are often employed for maneuvering and breaking~\citep{Walker1997Labriform}, station holding~\citep{arnold1991role} and stabilization~\citep{LauderDrucker2004}. Control surfaces located posterior to the center of mass produce trimming forces~\citep{webb1997designs,Webb2002,WilgaLauder2000_3Dkinematics} that can reorient the body and/or flow field passively, creating self-correcting forces that are stabilizing by nature, and are commonly used by fish swimming in steady and turbulent flows~\citep{Liao2007Review,webb1998Entrainment,pavlov2000patterns,Bartol2005boxfish}.Conversely, control surfaces anterior to the center of mass produce self-amplifying disturbances that promote higher maneuverability instead of stability~\citep{LauderDrucker2004,drucker2003function}. 

Unconventional underwater vehicle (UV) designs take inspiration in fish morphology and biomechanics to achieve high maneuverability. Flexible fish-like robots that can bend their bodies achieve high turning rates and small turning radii~\citep{Ay2018electronics,Hu2006design}, while improving the hydrodynamic performance of rectilinear motion~\citep{White_2021} compared to a rigid-body architecture~\citep{zhu2019tuna}. At the same time, flexible-body UVs typically display a complex internal architecture and a high number of moving parts. Bio-inspired, rigid-body UVs, on the other hand, benefit from a reduced number of moving parts that yields high electromechanical efficiency, as well as higher payload capacities compared to flexible-body designs~\citep{Liang2011twojoint,ko2025blueguppy}. The trend of reduced maneuverability of rigid-body UVs relative to flexible-body UVs can be addressed through the use of appendages as control surfaces~\citep{bandyopadhyay2002maneuvering,kato1998control,berlinger2017robust}, particularly at low forward speeds for small radius maneuvers and precise station holding, following a trend observed in nature \textemdash stiff-bodied fish tend to posses a greater number of appendages compared to flexible-bodied species~\citep{WEBB2005} that are used for maneuvering. As the three-dimensional kinematics and wake structures were revealed for pectoral fins of real fish~\citep{Drucker2002experimental,drucker2002wake,lauder2006locomotion}, computational simulations have been performed for finite aspect-ratio fins undergoing three-dimensional kinematics~\citep{beal2007harmonic,Bozkurttas2009,Dong2010computational}, and efforts in robotics typically focused on the development of sophisticated bio-inspired mechanisms with rigid and flexible fins~\citep{kato1998control,Kato2000Control,Suzuki2008load,Tangorra2010}. While greatly informative, it is difficult to generalize the hydrodynamic forces and flow structures from these results due to the high complexity of their fin shape, kinematics and in some cases the fin flexibility. 

In an effort to understand the the most salient parameters that dictate the force production of a flapping fin near a solid body and its ability to maneuver an underwater vehicle, we perform experiments of a simplified, bio-inspired flapping fin with a single degree-of-freedom. We use a model system to characterize the dynamics of a rigid underwater vehicle capable of lateral maneuvering and hovering through the use of control surfaces inspired by fish pectoral fins anterior to the vehicle's center of mass. For simplicity, our model is nominally two-dimensional with a fish-like profile and the fins are modeled as a pair of rigid flapping plates, attached to the sides of the vehicle as shown in Figures~\ref{fig:expSetup}(a) and~\ref{fig:expSetup}(c). The flapping plates thus impart momentum to the surrounding flow and move independently from the vehicle's body, generating powered forces~\citep{webb1997designs,Webb2002} to maneuver the vehicle.  

The system of a thin pitching plate attached to a solid boundary draws a parallel to the canonical problem of oscillating flapping foils. Depending on their kinematics, they can add momentum to the flow, producing a net thrust for propulsion~\citep{Floryan2017,moored2019inviscid} or extract momentum from the flow, producing a net drag for energy harvesting~\citep{Handy-Cardenas_2025,ribeiro2021wake,Kim2017energy}, for which the thrust or drag force production is associated with reverse von-K\`arm\`an and von-K\`arm\`an wakes, respectively~\citep{Schnipper2009}. In the case of self-propelling oscillating foils, the presence of a nearby solid boundary deflects the wake advection trajectory away from the boundary and affects its hydrodynamic performance; under ground-effect, the thrust is inversely proportional to the square of the distance to the boundary, increasing with smaller distances as the added-mass forces grow due to the presence of the boundary~\citep{Mivehchi2021,han2023revealing}. Unlike for those systems, however, the thin flapping plates studied here have their leading edge attached to the boundary, and the fin motion is not symmetrical relative to the freestream direction. This asymmetry introduces the net lateral forces that will be be explored for lateral maneuvering of the vehicle. Flow visualizations from~\cite{HeBreuer2026,he2023hydrodynamic} shows the formation of trailing-edge (TE) vortices during the fin's upstroke that eventually break-up during the downstroke. The advection trajectory of the TE vortex is shown to be affected by the fin kinematics. For low Strouhal, the TE vortex advects roughly along the streamwise direction. As the Strouhal number increases, the TE vortex trajectory becomes oblique, being accelerated in lateral direction. This was associated with the increased lateral force production with increasing flapping frequency and amplitude, and the peak in the lift amplitude throughout the flapping cycle roughly coinciding with the TE vortex shedding event. For $St\gtrsim0.4$, they observed an increased flow ``suction'' opposite to the streamwise direction in-between the fin and the body that produced vorticity with an opposite sign from the TE vortex that engulfs the TE vortex during the fin's downstroke as it advects downstream. 

Here we measure the streamwise and lateral forces produced by a single flapping plate and describe how it scales with the flapping kinematics. We then describe the forces produced by two plates flapping simultaneously for both a symmetric and anti-symmetric motion. Finally, a one-dimensional cyber-physical closed-loop control system~\citep{erickson2026vibrissa,zhu2021nonlinear,Onoue2015CPS,HOVER_TECHET_TRIANTAFYLLOU_1998} is employed to allow the vehicle to  move freely in the cross-stream direction in response to the lateral forces produced by the flapping plates.  

\section{Materials and Methods}

\begin{figure}[h]
    \centering
    \includegraphics[width=\linewidth]{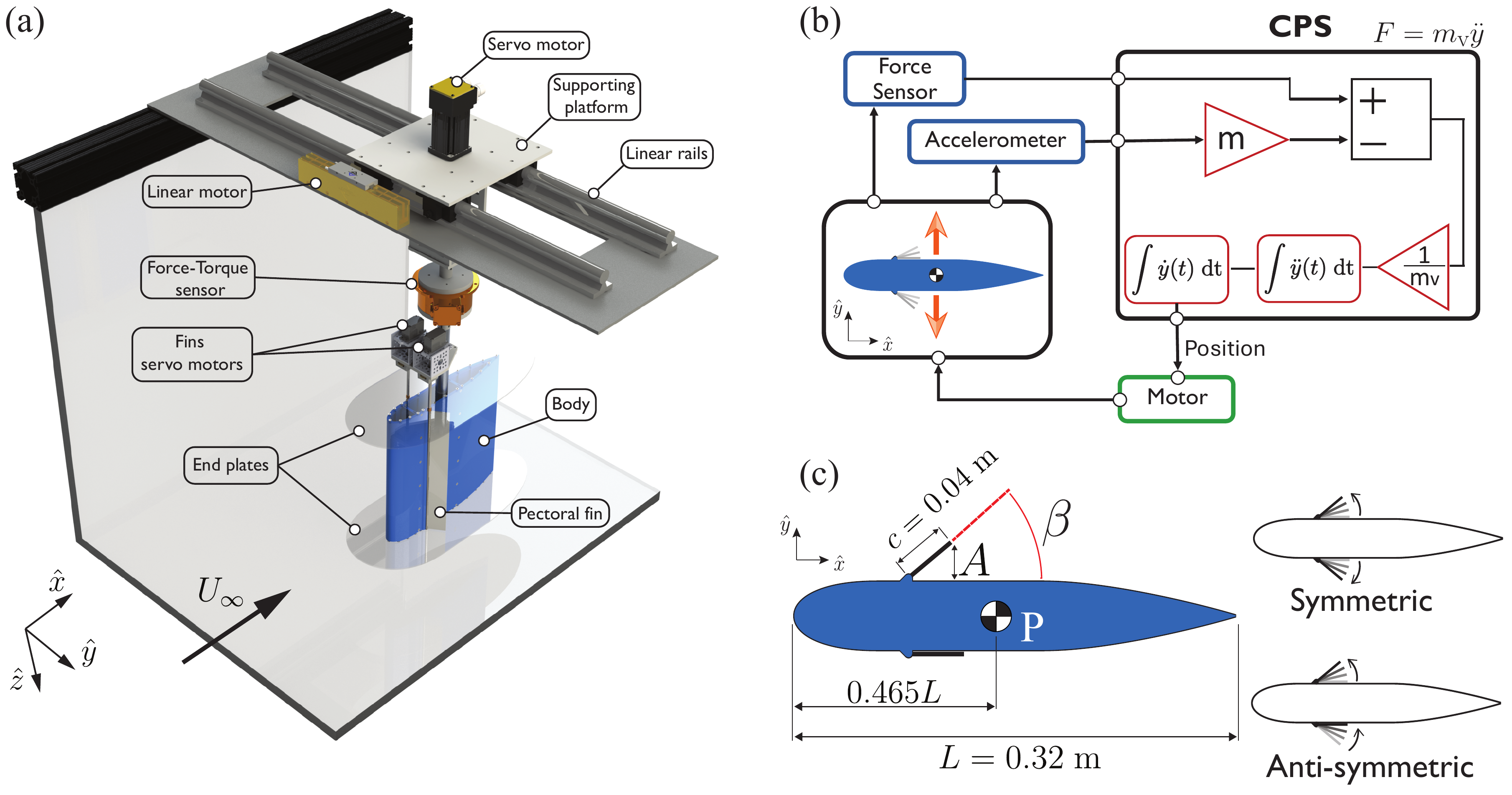}
    \caption{Experimental setup. (a) Schematics of the experimental apparatus. (b) Cyber-physical system control loop. (c) Left: Schematics of the vehicle profile and dimensions. Maximum fin amplitude $\beta$ is defined relative to the lateral body line. Right: Symmetric and anti-symmetric modes of two-fin flapping motions. }
    \label{fig:expSetup}
\end{figure}

Experiments were carried in a recirculating water channel with test section 4 m long, 0.8 m wide and 0.6 m deep. Figure~\ref{fig:expSetup}(a) shows the experimental setup. The two-dimensional fish-inspired model \citep{he2023hydrodynamic} has a main body (in blue) of length $L=0.32$ m and is designed as a modified NACA 0025 airfoil with constant thickness section inserted between $x/L = 0.19$ and $0.56$. End plates are placed at the top and bottom of the body. The ``pectoral fins" consist of a pair of rectangular flapping plates of chord length $c=0.04$ m and thickness $t_{\text{fin}}=3.2$\ mm, mounted on each side of the body with pitching axis located at $0.26 L$. The fins are each independently controlled by a servo motor (BLDC Planetary GearMotor, Anaheim Automation, model BLWRPG173D-24V-4000-R15-1000DN8).

A supporting shaft connects the fish model to a six-axis force and torque sensor (ATI Omega) that measures the instantaneous forces acting at point $P=(0.465L,0)$ relative to the vehicle's leading edge, as depicted in Figure~\ref{fig:expSetup}(c). The model is mounted to a supporting platform that can move in the cross-stream, $\hat{\text{y}}$-direction commanded by a linear motor. The model motion is governed by a real-time, one degree of freedom closed-loop cyber-Physical control system (CPS)~\citep{erickson2026vibrissa} (figure~\ref{fig:expSetup}(b)) which mimics a swimming fish that is free to move in the cross-stream direction in response to the cross-stream hydrodynamic forces, and subject to user-defined \emph{virtual} mass, $m_v$ (which can be different from the physical mass of the model, $m_p$). In these experiments the ratio between the virtual mass and the mass of water displaced by the submerged vehicle $m_f$ is $m^*=m_v/m_f=7.9$. The fluid force acting on the vehicle is defined as $\mathbf{F}_{\text{fluid}}(t) = \mathbf{F}_{\text{FT}}(t) - m_p a(t)$ where $\mathbf{F}_{\text{FT}}(t)$ is the cross-stream force measured by the force sensor, and $a(t)$ is the carriage acceleration measured by the accelerometer. These signals are then divided by a user-specified mass $m_v$ and integrated twice in time to determine the linear motor target position $y(t) = \iint \left( F_{\text{fluid}}(\tau)/m_v \right) \mathrm{d}\tau \mathrm{d}t$.

In this work the pectoral fin kinematics are a prescribed sinusoidal pitching motion. For simultaneous flapping of the two fins their phase synchrony is either symmetric or anti-symmetric as depicted in figure~\ref{fig:expSetup}(c). The angle of the right and left fins are $\theta_{R}(t) = \beta/2 \left( \sin{\left(2\pi f t\right)} + 1\right)$ and $\theta_{L}(t) = \beta/2 \left( \sin{\left(2\pi f t - \phi\right)} + 1\right)$, respectively, where $f$ is the fin frequency (Hz), $\beta$ is the maximum fin amplitude measured from the side of the body, and the phase synchrony is $\phi = \pi$ for a symmetric motion or $\phi = 0$ for an anti-symmetric motion. The reduced frequency is defined as $k = fc/U$, and the Strouhal number is $St=fA/U$, where $A = c \sin \beta$ is the projected frontal area of the fin measured at the peak amplitude of the fin in the cross-stream direction as indicated in figure~\ref{fig:expSetup}(c). A constant flow speed $U=0.2$ m/s is imposed throughout this study, with Reynolds number based on the body length $Re_{L} = 60\times 10^3$ and chord-based Reynolds $Re_c=7.6 \times 10^3$. 

A quasi-steady fin regime is also measured for comparison to the unsteady, flapping case defined above. In the quasi-steady regime, the fin moves slowly from its retracted position $\theta_0=0^{\circ}$ to a maximum angle $\theta_{max}=100^{\circ}$, and then back to $\theta = 0^{\circ}$ at a constant rate of $\pi/90 \text{(rad/s)}=2(^{\circ}\text{/s})$, yielding a reduced frequency $k=0.002$. At such a low frequency, any unsteadiness due to the fin motion is negligible, and the forces measured against the fin angle are treated as quasi-steady.

The non-dimensional forces and  $\hat{z}$-moment measured at point $P=(0.465L,0)$ of the vehicle are defined as
\begin{equation}
    C_{F_{x}} = \frac{F_x}{q_{dyn}c} \quad {,} \quad C_{F_{y}} = \frac{F_y}{q_{dyn}c}\quad {and} \quad C_{M_{z}}=\frac{M_z}{q_{dyn} c^2},
\end{equation}
respectively, where$F_x$  and $F_y$ are the forces in the streamwise and cross-stream directions, respectively, $M_z$ is the moment,  $q_{dyn} = 0.5\rho U^2$ is the dynamic pressure based on the freestream speed $U$, and $c$ is the fin cord-length. For two fins flapping simultaneously, $F_x = F_x^R + F_x^L$, $F_y = F_y^R+F_y^L$ and $M_z=M_z^R+M_z^L$ are the net forces experienced by the body produced by the pair for a  symmetric or anti-symmetric motion where the superscripts $()^R$ and $()^L$ denote the right- and left-side fins, respectively. The reported coefficients for forces and moment are averaged over three trials, each calculated from a minimum of 20 measured flapping cycles. The standard error, $SE = \sigma/\sqrt{N}$, is computed for all measurements, where $\sigma$ is the sample standard deviation and $N$ is the number of trials. It is visualized as shaded regions whenever it exceeds the symbol size. If no shaded region is visible, the standard error $SE$ is smaller than the plotted symbols. As depicted in Figures~\ref{fig:expSetup}(a) and~\ref{fig:expSetup}(b), the streamwise $\hat{x}$-axis is positive in the direction and sense of the free stream, thereby defining drag as a positive streamwise force (towards the rear of the model). Conversely, thrust is a negative streamwise force.

\section{Results}

\subsection{Characterization of fin forces and $z$-moment}

\begin{figure}[h]
    \centering
    \includegraphics[width=\linewidth]{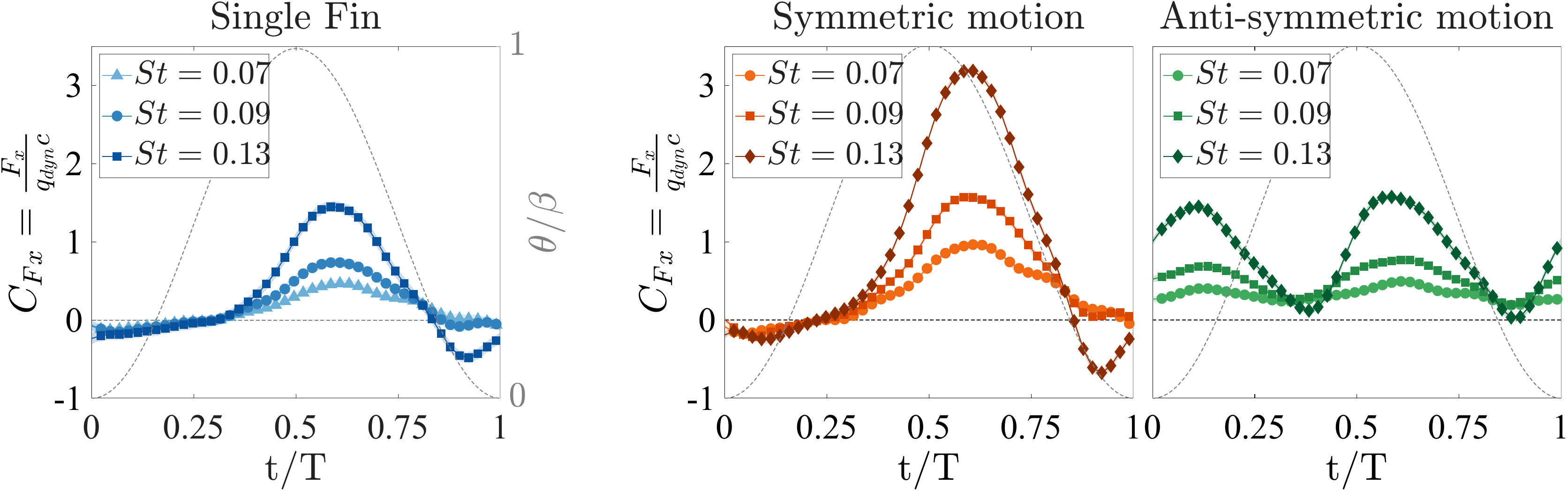}
    \caption{Phase-averaged, streamwise forces produced by (a) a single fin, and (b) and (c) two fins flapping with phase synchronization $\phi=\pi$ and $\phi=0$, respectively. Fin amplitudes $[30^{\circ},\  40^{\circ},\ 60^{\circ}]$ and frequency is fixed at $f=0.75$ Hz and $k=0.15$. The shaded region indicates the standard error interval for a total of three measurements. The gray dashed line shows the fin angle over one cycle $\theta/\beta$ as indicated by the vertical axis on the right-hand side. Positive values correspond to drag, while negative values correspond to thrust.}   \label{fig:CFx}
\end{figure}

The streamwise and lateral forces produced by a single fin flapping are measured for a vehicle subject to an oncoming flow $U = 0.2$ m/s for a fixed flapping frequency $f=0.75 Hz$, reduced frequency $k=0.15$, and varying amplitudes $\beta = [30^{\circ},  40^{\circ}, 60^{\circ}]$ and associated Strouhal numbers $St= [0.07,\ 0.09,\ 0.13]$. The phase-averaged, streamwise force coefficient $C_{F_x}$ produced by a single flapping fin is shown in figure~\ref{fig:CFx}(a). A positive streamwise force $C_{F_{x}}>0$ denotes drag, whereas a negative streamwise force $C_{F_{x}}<0$ denotes thrust. A positive force $C_{F_x}>0$ is measured for all Strouhal cases during the interval $0.25<t/T<0.75$, for which the instantaneous angle of the fin $\theta$ is large. The peak in the force magnitude increases monotonically with Strouhal. During the downstroke motion, a small thrust force is produced by the highest Strouhal case only. This is associated with flow ejected from the region between the fin and the side wall of the vehicle~\citep{drucker2000hydrodynamic,he2023hydrodynamic,ormonde2024positional}, imparting downstream momentum to the fluid, which produces a force with an upstream component andwhich is responsible for the thrust force observed at the end of the downstroke motion.

Figures~\ref{fig:CFx}(b) and (c) show the measured streamwise force from the simultaneous flapping of two fins for symmetric and anti-symmetric synchronization, respectively. For a symmetric motion with $\phi = \pi$, the up- and downstroke of both fins coincide, leading to the phase-averaged drag force profile to be equal to that of a single fin, but twice as strong. For an anti-symmetric motion $\phi=0$, on the other hand, lower variations in the force are observed when compared to the symmetric motion case as shown in figure~\ref{fig:CFx}(c), as the peaks from each fin do not overlap since when one fin is fully open the opposite-side fin is fully retracted and produces close to zero force.

\begin{figure}
    \centering
    \includegraphics[width=\linewidth]{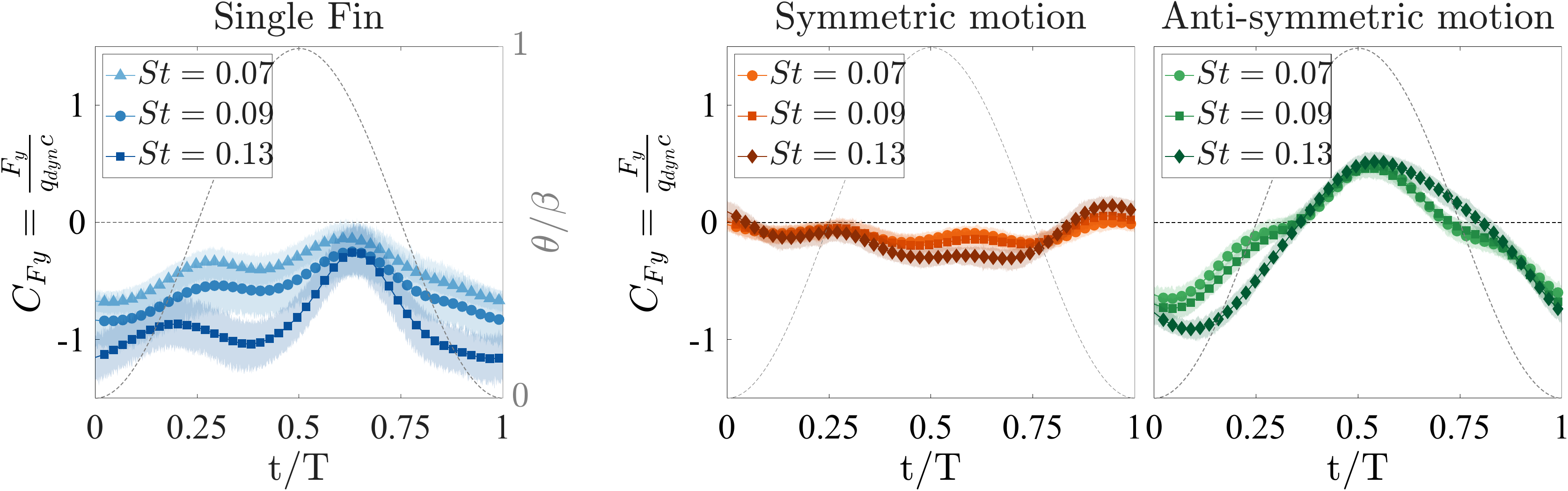}
    \caption{Phase-averaged, cross-stream forces produced by (a) a single fin, and (b) and (c) two fins flapping with phase synchronization $\phi=\pi$ and $\phi=0$, respectively. Fin amplitudes $\beta = [30^{\circ},\  40^{\circ},\ 60^{\circ}]$ and frequency is fixed at $f=0.75$ Hz and $k=0.15$. The shaded region indicates the standard error interval ($SE$) for a total of three measurements. The gray dashed line indicates the fin angle over one cycle $\theta/\beta$ as indicated by the vertical axis on the right-hand side.}
    \label{fig:CFy}
\end{figure}

The phase-averaged lateral force produced by the right-side fin is shown in figure~\ref{fig:CFy}(a). A negative time-averaged lateral force is observed for all Strouhal numbers, i.e., the force points in the direction opposite to the side on which the flapping fin is located. Increasing the Strouhal number leads to higher amplitude oscillations of the lateral force over one flapping cycle, similarly to what is observed for the streamwise force. Figures~\ref{fig:CFy}(b) and (c) show the streamwise force for two fins flapping with symmetric ($\phi=\pi$) and anti-symmetric ($\phi=0$) motion, respectively. For both synchronies the time-average lateral force is effectively zero, and the instantaneous forces are suppressed throughout the entire flapping cycle for the symmetric motion due to phase cancellation of the left- and right-side fins. Looking at the streamwise (drag) and cross-stream (lift) force profiles for a single flapping fin in figures~\ref{fig:CFx}(a) and~\ref{fig:CFy}(a), respectively, it can be observed that, for all three Strouhal numbers the peak in the time-averaged drag magnitude coincides with the minimum in the lift force.

The force data shows how the fin generates forces in the horizontal, $x-y$ plane, and simultaneous flapping of two fins can be leveraged to amplify or suppress the streamwise and cross-stream components of the forces. The symmetric motion $\phi=\pi$, for example, effectively eliminates the lateral forces while preserving the profile of the phase-averaged streamwise force generation. It can thus be employed for fast decelerations of the vehicle, for example, without compromising its attitude relative to the swimming direction. An anti-symmetric motion, on the other hand, decreases the intra-cycle fluctuations of the streamwise force component while retaining the oscillatory behavior of the lateral force observed for a single fin. Since the fins are each located at one side of the vehicle, the flow structures produced by each fin do not interact among themselves, and their hydrodynamic forces are independent from one another. For symmetric fin motion, the streamwise forces produced by the right and left side fins are the same $F_x^R(t/T) = F_x^L(t/T)$, while the lateral forces have opposite signs $F_y^R(t/T) = - F_y^L(t/T)$.

\begin{figure}
    \centering
    \includegraphics[width=\linewidth]{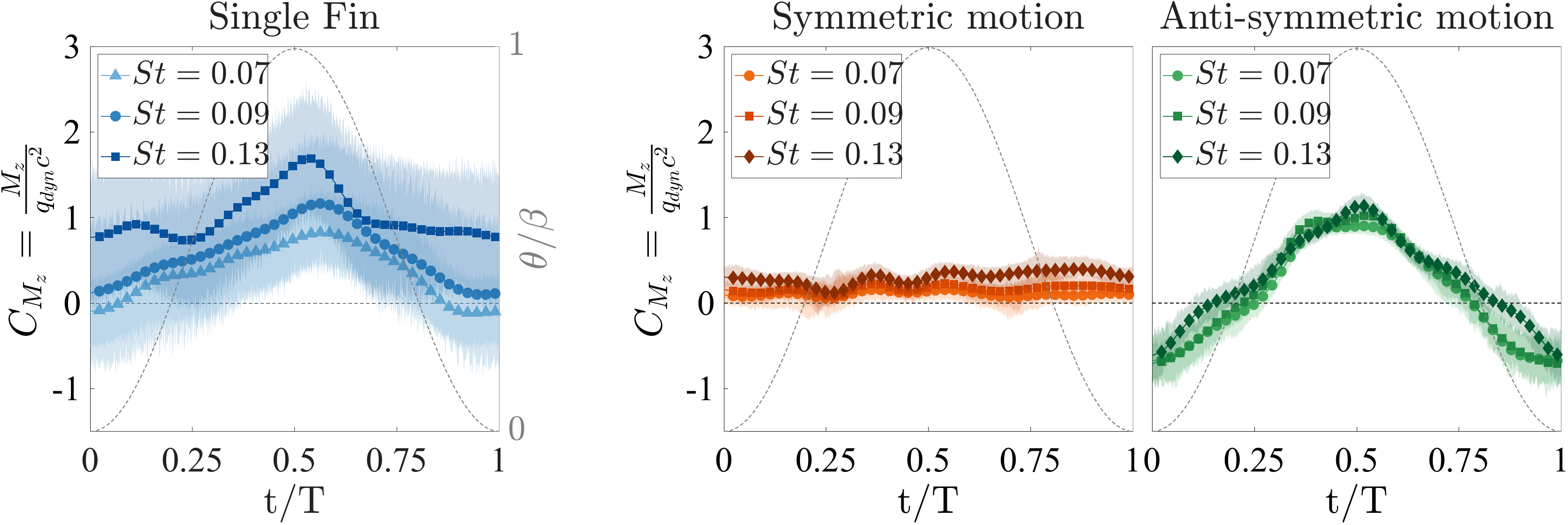}
    \caption{Phase-averaged $\hat{z}-$moment produced by (a) a single fin, and (b) and (c) two fins flapping with phase synchronization $\phi=\pi$ and $\phi=0$, respectively. Amplitudes $ \beta=[30^{\circ},\  40^{\circ},\ 60^{\circ}]$ and frequency $f=0.75$ Hz and $k=0.15$. The shaded region indicates the standard error interval ($SE$) for a total of three measurements. The gray dashed line indicates the fin angle over one cycle $\theta/\beta$ as indicated by the vertical axis on the right-hand side of figure (a).}
    \label{fig:CMz}
\end{figure}

Figure~\ref{fig:CMz} (a) shows the measured moment coefficient for a single fin. Positive (clockwise) moments with peaks at instant $t/T\approx0.56$ are observed for all three frequencies. For the cases with $St=0.09$ and $0.07$ the moment force starts from zero, monotonically increasing until reaching its peak and then returning to zero at the end of the flapping cycle. For the higher amplitude case (square markers) on the other hand, the moment does not return to zero at the end of each cycle, achieving a minimum value of about 0.8, which is almost as high as the peak moment produced by the lower Strouhal cases (circle and triangle markers), and thus distinctly different. Figures~\ref{fig:CMz}(b) and (c) show the $\hat{z}-$moment for symmetric and anti-symmetric motion, respectively. The moment signal profile is suppressed throughout the entire flapping cycle for a symmetric motion, whereas the anti-symmetric case shows a similar behavior to the lift for the same synchronization. These results suggest that the main component that contributes to the $\hat{z}-$moment is the cross-stream (lift) component, and not the streamwise (drag) component of the force. 

The measured standard error for the cross-stream force (Figure~\ref{fig:CFy}) and the $\hat{z}$-moment (Figure~\ref{fig:CMz}) signals is significantly larger than that of the streamwise force (Figure~\ref{fig:CFx}). We attribute this discrepancy to structural vibrations of the experimental apparatus. The streamlined profile of the vehicle leads to a small displacement of fluid in the streamwise direction in response to the oscillatory forces produced by the flapping fins. Conversely, the large projected area of the vehicle in the vertical $\hat{x}-\hat{z}$ plane displaces a large volume of fluid, resulting in a large added mass force~\citep{brennen1982review} that is captured by the force and torque transducer. That said, the frequency of the structural oscillations is not phase-locked to the flapping frequency, thus the phase-averaging of the raw signals successfully rejects the structural vibration noise from the hydrodynamic forces of interest, which allows us to analyze the phase-averaged force and moment data without any additional filtering. 

\subsection{Scaling of the streamwise and cross-stream forces}
Experiments with a single flapping fin are performed for a range of frequencies $f = [0.5, 0.6, ..., 1.2]$ Hz, and amplitudes $\beta=[20^{\circ}, 30^{\circ},40^{\circ}, 50^{\circ}]$ at fixed free-stream speed $U=0.20$ m/s and chord-based Reynolds number $Re_c = 7.6\times10^3$.  

\begin{figure}
    \centering
    \includegraphics[width=0.8\linewidth]{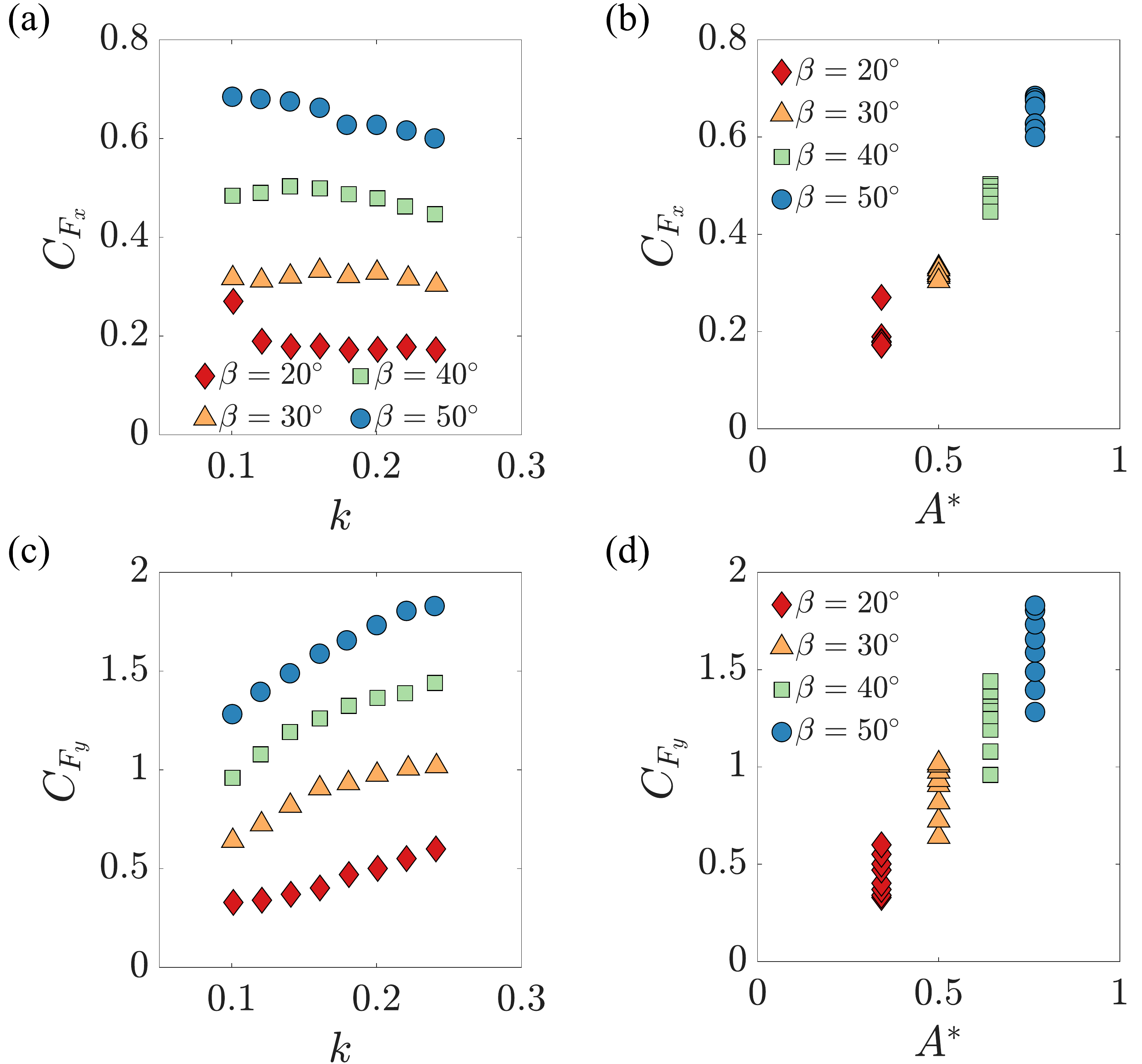}
    \caption{Time-average forces produced by a single flapping fin for a stationary vehicle for input parameters $f = [0.5, 0.6, ..., 1.2]$ Hz and $\beta = [20^{\circ}, 30^{\circ}, 40^{\circ}, 50^{\circ}]$. (a) Streamwise (drag) force coefficient vs. reduced frequency $k=f c / U$ for different amplitudes $\beta$. The force is primarily a function of the pitching amplitude, with a weak dependence on the frequency. (b) Streamwise (drag) force coefficient vs. non-dimensional amplitude $A^*=\sin{\beta}$. The streamwise (drag) force is approximately a linear function of amplitude, with minor variations for the range of frequencies tested. (c) Cross-stream (lateral) force vs. reduced frequency $k$ for different amplitudes $\beta$. (d) Cross-stream (lateral) force vs. non-dimensional amplitude $A^*$. }
    \label{fig:CF_scaling}
\end{figure}

The time-average, streamwise forces $C_{F{x}}$ are presented in figures~\ref{fig:CF_scaling}(a) and~\ref{fig:CF_scaling}(b) as a function of reduced frequency $k=fc/U$ and amplitude $A^*=\sin{\beta}$, respectively. To first order, the streamwise (drag) force is insensitive to changes in the frequency $k$ (fig.~\ref{fig:CF_scaling}(a)), with a slight decrease in the force magnitude observed for the two highest amplitudes $\beta = 40^{\circ}$ (green squares) and $50^\circ$ (blue circles). The streamwise force is much more sensitive to the flapping amplitude $A^* = \sin{\beta}$. Figure~\ref{fig:CF_scaling}(b) shows a large variation in the force coefficient for the amplitudes tested, going from an average $\overline{C}_{F_x}= 0.19$ for $\beta = 20^{\circ}$ (red diamonds) to $\overline{C}_{F_x} = 0.65$ for the highest amplitude $\beta = 50^{\circ}$. The plot shows an approximately linear relationship between the streamwise (drag) force coefficient and the projected frontal area $A^* = \sin{\beta}$.

\begin{figure}
    \centering
    \includegraphics[width=\linewidth]{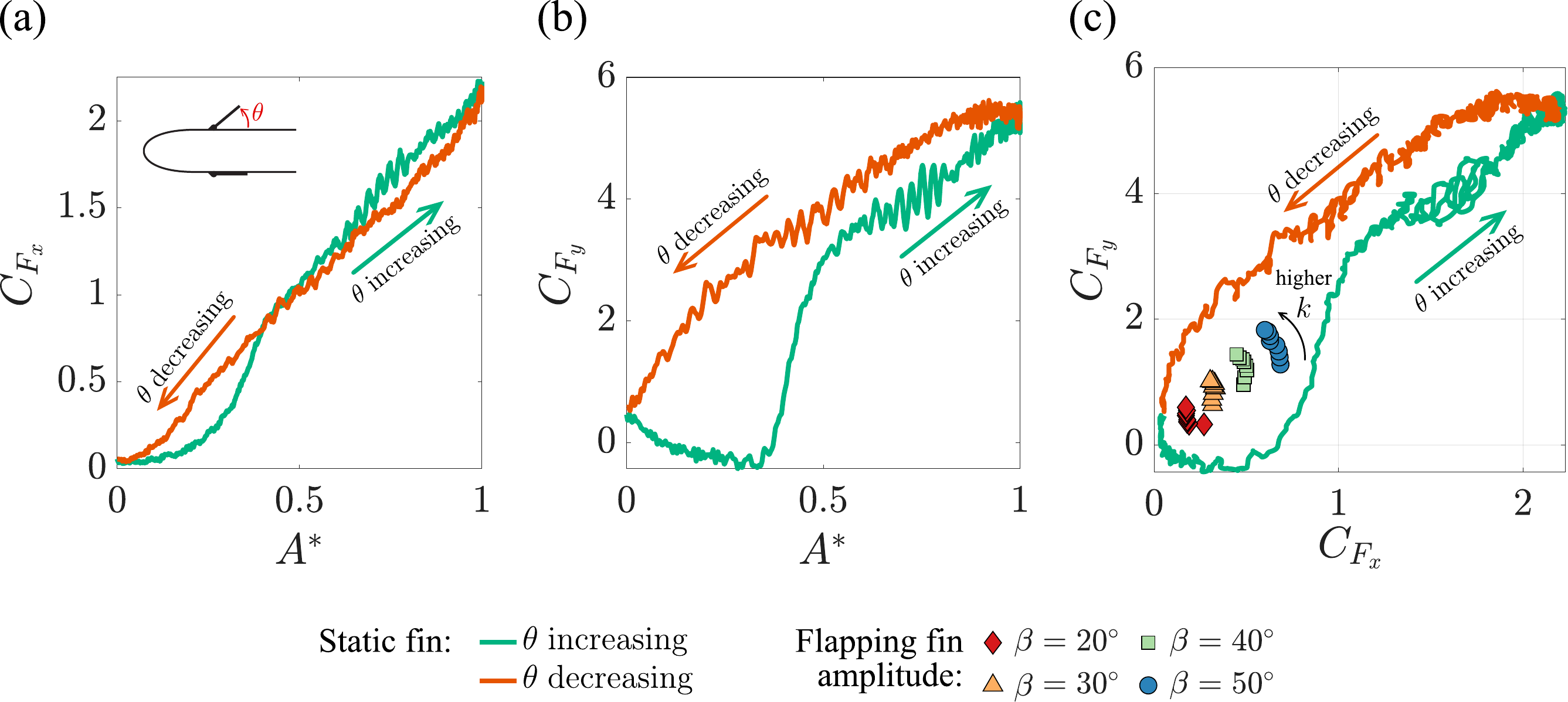}
    \caption{Average forces produced by a quasi-steady fin during an upstroke motion (green line) starting from a retracted position $\theta_0=0^{\circ}$ to the maximum angle $\theta_{max}=100^{\circ}$, and a downstroke motion (orange line) back to $\theta=0^{\circ}$. (a): Streamwise (drag) force vs. amplitude $A^*=\sin{\theta}$. $C_{F_x}$ is a linear function of $A^*$, except at low amplitudes for an upstroke motion (green line). (b): Cross-stream force $C_{F_x}$ vs. amplitude $A^*$. Large hysteresis in the lateral force is caused by non-linear flow attachment during the initial phase of the upstroke motion compared to a fully detached flow during the downstroke. (c) $C_{F_y}$ vs. $C_{F_x}$ for the quasi-steady fin (lines) compared to the time-averaged coefficients for a flapping fin (markers). The flapping fin forces seem bounded by the quasi-steady curve, and the lateral to streamwise force ratio increases with the flapping frequency $k=fc/U$.}
    \label{fig:staticForces}
\end{figure}

The streamwise force coefficient $C_{F_x}$ presented on Figure~\ref{fig:CF_scaling}(b) can be discussed in light of the forces measured for a quasi-steady case, for which the fin is slowly moved from its retracted position $\theta_0=0^{\circ}$ to a maximum angle $\theta_{max}=100^{\circ}$, and then back to $\theta = 0^{\circ}$. Due to the extremely small reduced frequency of the experiment $k=0.02$, the unsteady effects from the fin motion are negligible. Figure~\ref{fig:staticForces}(a) presents the quasi-steady, streamwise force data as a function of $A^*=\sin{\theta}$. The green line shows the force for an increasing fin angle up to $\theta_{max}$, whereas the orange line indicates the measured force for a decreasing fin angle. The quasi-steady, streamwise (drag) force is a linear function of $A^*$, except for the region $0\leq A^*<0.4$ when the fin is moving away from the body starting from the retracted position 
$\theta_0=0^{\circ}$. This linear behavior indicates that the streamwise force can be interpreted as a pressure-based drag, typically observed for bluff bodies at sufficiently high Reynolds numbers~\citep{roshko1955wake}. It is important to note that the definition of $A^*=\sin{\theta}$ for the quasi-steady case represents the instantaneous, projected length of the fin in the cross-stream direction, whereas for the flapping fin case $A^*$ is defined by the maximum fin amplitude $\beta$ as the fin angle varies $0\leq\theta\leq\beta$ throughout the flapping cycle. For that reason, different values of the streamwise force coefficient are expected between the two cases, and neither of the two datasets are universal definitions of the fin's drag coefficient since they represent two different regimes \textemdash the unsteady flapping case of Figure~\ref{fig:CF_scaling}(b) and the quasi-steady case of Figure~\ref{fig:staticForces}. That said, what \emph{can} be inferred from both cases is that, to first order, the streamwise force is proportional to the projected frontal length (or area) of the fin, and is largely insensitive to the flapping frequency for the unsteady case within the range of reduced frequencies and Strouhal numbers studied here.  

Figure~\ref{fig:staticForces}(b) presents the cross-stream force coefficient $C_{F_y}$ vs. $A^*$ for the quasi-steady case. During the upstroke, as the fin angle increases from $\theta_0=0$ (green line), the cross-stream force increases abruptly at $A^*=0.4$ (or $\theta = 24^{\circ}$), then increasing monotonically afterwards. During the downstroke (orange line), however, the force decreases monotonically for the entire range from $\theta_{max}$ to $\theta_0$. This hysteretic behavior is attributed to the shear layer produced at the fin's leading edge, which, during the upstroke motion starting from $\theta_0$, stays continuously attached to the fin and the aft portion of vehicle, only ``breaking up'' at sufficiently high angles of attack. Once this attached shear layer breaks down, a pronounced vortex shedding occurs at the fin's trailing edge, and the flow is only able to reattach to the aft portion of the body when the fin is fully retracted again. This non-linear flow behavior explains the dramatic differences in the lateral force production between the upstroke (green line) and downstroke (orange line). A detailed description of this phenomenon, along with flow field visualizations is found on~\citep{he2023hydrodynamic}.

Figure~\ref{fig:staticForces}(c) shows the coefficients of cross-stream force vs. streamwise force for both the quasi-steady case (solid lines) and the flapping case (markers). For any given value of streamwise (drag) force $C_{F_x}$, the plot clearly shows that the lateral force $C_{F_y}$ is always larger for the downstroke (orange line) than for the upstroke (green line). This indicates that the flow structures along the fin and body dramatically affect the forces produced by the fin, and the $C_{F_y}/C_{F_x}$ ratio seems to be higher for the fully detached flow observed by~\cite{he2023hydrodynamic,HeBreuer2026} during the downstroke (orange line). The forces from a flapping fin are also plotted on Figure~\ref{fig:staticForces} for comparison (markers). The plot shows how both forces increase with larger flapping amplitudes $\beta$, as previously seen on Figure~\ref{fig:CF_scaling}, and also helps visualize that higher flapping frequencies lead to an increase in the $C_{F_y}/C_{F_x}$ ratio. Interestingly, all flapping cases are bounded by the two lines produced by the quasi-steady fin, suggesting that the maximum achievable $C_{F_y}/C_{F_x}$ ratio occurs for a static fin when the flow field is fully detached downstream of the fin's trailing edge.

\begin{figure}
    \centering
    \includegraphics[width=0.8\linewidth]{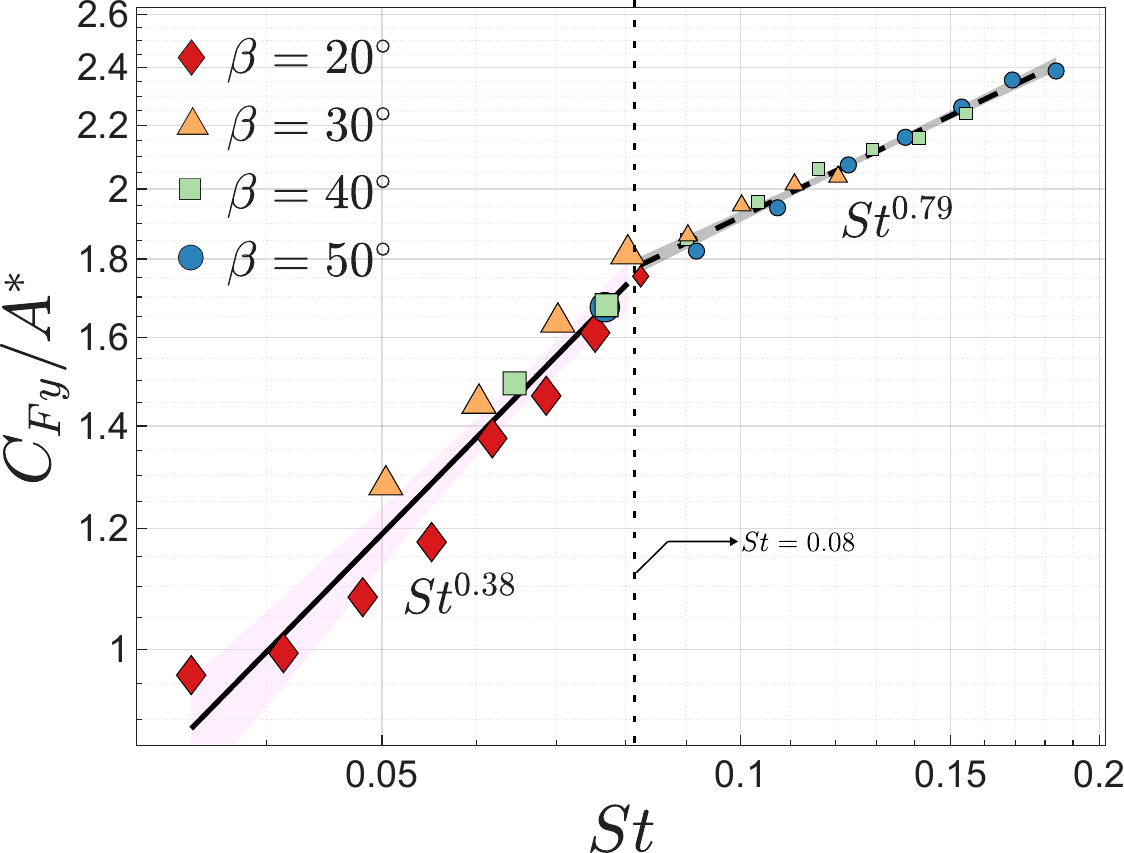}
    \caption{Lateral force coefficient divided by the projected cross-stream length $C_{Fy}/A^*$ as a function of Strouhal number $St$. Data points are shown as colored symbols, and dashed-lines indicate power laws of the type $St^n$, where $n=1.0$, 0.5 and 0.4 as indicated by the labels next to the lines. For the higher amplitude cases tested $\beta=30^{\circ}, 40^{\circ}$ and $50^{\circ}$, the data trend is approximately a straight line with the exponent of Strouhal number $n=0.4$ and $n=0.5$ } 
    
    \label{fig:CFy_scaling}
\end{figure}

The time-averaged lateral force produced by a single flapping fin varies with the flapping frequency and amplitude. Figure~\ref{fig:CFy_scaling} shows the cross-stream force coefficient divided by the projected cross-stream length $C_{Fy}/A^*$ as a function of the Strouhal number in a log-log plot as colored symbols. 

Since the Strouhal number $St$, reduced frequency $k$ and non-dimensional amplitude $A^*$ have been shown to be key parameters for scaling laws of thrust-producing oscillating plates and foils~\citep{smits2019undulatory,moored2019inviscid}, including for foils near a solid boundary~\citep{Quinn2014}, we use those same quantities as the building blocks for a data-driven scaling law. 
The data indicates that the lateral force follows not one, but two power laws of the form $C_{Fy}/A^* \propto St^\alpha$, where the value of the exponent $\alpha$ depends on the range of Strouhal number of the flapping fin. The transition in the lateral force scaling from one regime to another occurs at the crossover Strouhal number $St_{c}=0.08$, indicated by the thin vertical dashed line. For $St<St_c$, the data scales with $\alpha_1=0.38$ as indicated by the solid black line. For $St>St_c$, the data scales with $\alpha_2\approx2\alpha_1=0.79$ as indicated by the dashed black line. The shaded areas represent the 95\% confidence intervals for each linear regression. In order to determine $St_c$ and the two proposed power laws, the corrected Akaike information criterion (AICc) for small samples~\citep{StoicaSelen2004} was combined with iterative piecewise linear regression computations. First, the dataset was sorted in ascending order based on the independent variable, $St$. Every data point was then evaluated as a potential critical threshold, $St_c$. At each iteration, the data was split into two subsets: one containing values below and the other containing values above the trial threshold. Linear regressions were then fitted to both subsets, and the Akaike Information Criterion was calculated for each candidate piecewise regression against a single linear regression of the entire dataset. After testing every possible split point across the dataset, the final critical threshold, $St_c$, and its corresponding piecewise model was selected from the configuration that yielded the lowest AIC score, representing the optimal goodness-of-fit regression. A Bayesian information criterion (BIC)~\citep{KassRaftery1995} test was also performed on the optimal piecewise regression, resulting in a Bayes factor $2\ln B_{10}\approx \Delta BIC =24.4$, which is substantially higher then the $2\ln B_{10}>10$ threshold proposed by \citep{KassRaftery1995}, and provides very strong evidence in favor of the obtained piecewise regression. Finally, The 95\% confidence intervals for the two linear regressions were computed via bootstrap resampling~\citep{james2013introduction} with 1,000 resamples.

The proposed scaling can also be defined in terms of the flapping frequency and amplitude as 
\begin{align}
C_{Fy} &\propto A^*St^\alpha
\label{eqn:CFy_scaling_1} \\
C_{Fy} &\propto (\sin{\beta})^{1+\alpha} \, k^{\alpha}
\label{eqn:CFy_scaling_2}
\end{align}

Equations~\eqref{eqn:CFy_scaling_1} and~\eqref{eqn:CFy_scaling_2} are each comprised of two non-dimensional terms. The first term is a characteristic length, expressed as $A^*$ in equation~\eqref{eqn:CFy_scaling_1}, and $\sin{\beta}$ in equation~\eqref{eqn:CFy_scaling_2}. It can be interpreted as the projected frontal area of the fin at its maximum angle. The second term in equation~\eqref{eqn:CFy_scaling_1} is the Strouhal number $St=f A/U$, which is the ratio between two velocities: $f A$ is a reference velocity of the fin's trailing edge in the cross-stream direction, and $U$ is the reference velocity in the streamwise direction. In equation~\eqref{eqn:CFy_scaling_2}, the second term is the reduced frequency $k=f c/U$, which is a ratio between two time scales: $1/f$ is the period of the fin's oscillation, and $c/U$ is the time it takes for a fluid particle to traverse a distance of one chord-length at the reference velocity $U$. 

While the proposed scaling defined by Equations~\eqref{eqn:CFy_scaling_1} and~\eqref{eqn:CFy_scaling_2} collapses the data well (see Figure~\ref{fig:CFy_scaling}), it is worth noting that it is not derived from first principles, but rather through a physics-inspired data-driven procedure. That said, the  scaling for the streamwise forces (see Figure~\ref{fig:CF_scaling}(b)) and cross-stream forces (see Figure~\ref{fig:CFy_scaling} and equations~\eqref{eqn:CFy_scaling_1} and~\eqref{eqn:CFy_scaling_2}) are  capable of predicting the fin forces as a function of its kinematics, and have a practical application in the design and control of bio-inspired, flapping control surfaces for underwater vehicles. The piecewise scaling law obtained for $C_{Fy}$ indicates a change in the flapping fin regime, which we hypothesize is due to qualitative changes in the characteristics of the flow field and vortex dynamics shed at the fin's trailing edge \cite{HeBreuer2026}. Further flow visualization experimental data should help clarify the two regimes observed in the lateral force data described in Figure~\ref{fig:CFy_scaling}.  

\subsection{Lateral Maneuvering}
Next, we demonstrate how the pectoral fins are used to maneuver the ``fish'' in a bio-mimetic manner.  
 
The Cyber-physical system on which the test article is mounted allows the vehicle to translate in the cross-stream direction in response to the lateral forces generated by the fins, while maintaining the streamwise alignment i.e., at a zero yaw angle. 
 
\begin{figure}
    \centering
    \includegraphics[width=0.85\linewidth]{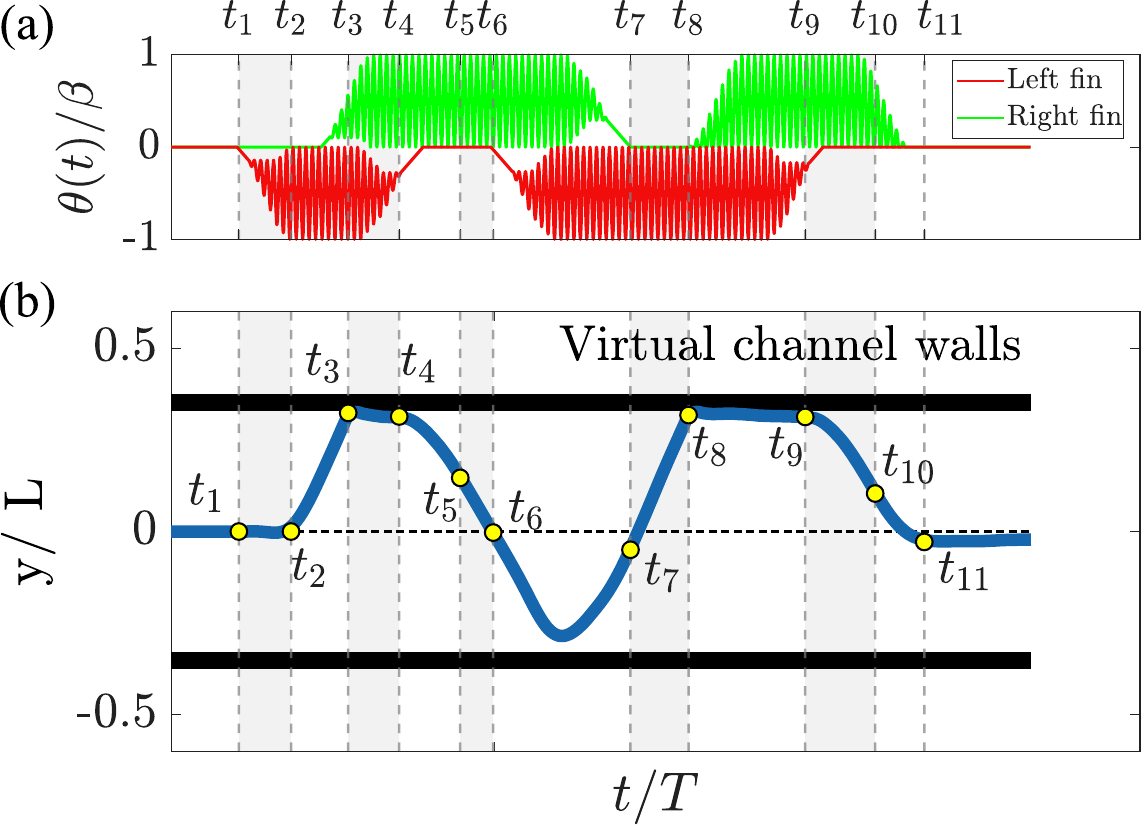}
    \caption{Lateral maneuvering of the vehicle in response to lateral forces produced by two flapping fins using the Cyber-physical system. (a) Angle $\theta/\beta$ of the left (red) and right (green) fins as a function of time $t/T$. Flapping frequency $f=1.25$Hz and amplitude $\beta=45^{\circ}$. (b) Lateral position of the vehicle in time. (c) Fluid force $F_f=F_{FT} - m_p\ddot{y}$ (black line) and inertial force $m_v\ddot{y}$ of the Cyber-physical vehicle in time.}
    \label{fig:CPSmotion}
\end{figure}

In this demonstration, we operate the two fins in a Left-Right-Left-Right sequence, as shown in Figure~\ref{fig:CPSmotion}(a), which shows the angular position of the flapping fins as a function of time. Figure~\ref{fig:CPSmotion}(b) shows the resulting position of the vehicle as a function of time. The cross-stream position of the vehicle is non-dimensionalized by its total length $L$. Virtual walls (black lines) are imposed by software to maintain the vehicle at a minimum distance of $Y/L$ of the water channel walls thus avoiding any near-boundary effects with the test facility. 

Initially at rest at $t/T=0$, the left fin (red line) starts flapping at $t_1$, generating an increasingly positive force $F_f>0$. At $t/T = t_2$ the fin achieves its total prescribed flapping amplitude and the vehicle's velocity increases quickly. From $t_2$ to $t_3$, the vehicle moves in the $+y$-direction, briefly achieving its terminal velocity $v_{eq}$ a few flapping cycles before instant $t_3$. At the terminal velocity, the vehicle's acceleration is zero and the net fluid force $F_f\approx0$ since the total cross-stream force produced by the fin is balanced by  opposing hydrodynamic forces due to its lateral velocity $v_{eq}/U=0.064\pm0.009$. At $t_3$, the carriage is forced to stop abruptly at the virtual wall (black thick line in Figure~\ref{fig:CPSmotion}(b)). From $t_3$ to $t_4$, the left fin (red line, Figure~\ref{fig:CPSmotion}(a)) is slowly retracted while the right fin (green line) starts flapping with increasing amplitude. At $t_4$, the fluid force becomes negative $F_f<0$ and the vehicle starts moving in the $-y$-direction. From $t_4$ to $t_5$, the vehicle's lateral speed increases. From $t_5$ to $t_6$ the vehicle travels at a roughly constant speed $v_{eq}$, as indicated by the straight line segment within those instants in Figure~\ref{fig:CPSmotion}(b). At instant $t_6$ the left fin is commanded to flap again, \emph{decelerating} the vehicle until it stops (without hitting the virtual wall).  The vehicle then accelerates in the $+y$-direction as the right fin (green line) is ramped down. From $t_7$ to $t_8$ only the left fin is flapping, and the vehicle travels at its terminal velocity again.  The vehicle is allowed to hit the virtual wall at $t_8$.  From $t_8$ to $t_9$ the right fin (green line) is commanded while the left fin (red line) is retracted.  Once the net fluid force acting on the vehicle becomes negative ($F_f<0$) at $t_9$, the vehicle accelerates in the $-y$-direction. The right fin is slowly retracted once more, and at $t_10$ the fin flapping amplitude is too small to overcome the hydrodynamic forces due to the vehicle's instantaneous lateral velocity and the vehicle starts to decelerate. Starting at $t_10$, as the right fin (green line) retracts the vehicle decelerates further, due to the hydrodynamic force caused by its lateral velocity. At $t_11$ the fins are fully retracted and the vehicle's velocity and acceleration are very small. At that point, $F_f\approx0$ and $v\approx0$, i.e, the vehicle maintains a constant position until the end of the experiment.

\section{Discussion and Conclusions}

In this study, flapping plates attached to the sides of an underwater vehicle are used as bio-inspired control surfaces. The forces produced by the fins were characterized as a function of its kinematics for a single fin and for the simultaneous flapping of both fins with symmetric and anti-symmetric synchronizations. A symmetric motion shows an almost perfect suppression of the instantaneous lateral force and $\hat{z}$-moment due to phase cancellation, while the instantaneous streamwise force production for two fins is just twice that of a single fin. A symmetric motion is thus suitable for fast deceleration maneuvers in the streamwise direction without disturbances in the lateral direction or yaw angle of a vehicle. Conversely, non-symmetric motions produce non-zero lateral forces and $\hat{z}$-moments within one flapping cycle. These forces and moment may be exploited for fine control of a vehicle's yaw angle, as well as to suppress transient flow disturbances that can disturb the vehicle's orientation relative to its traveling direction.

We find that the streamwise and cross-stream forces produced by the flapping fins are strongly dependent on the flapping amplitude and frequency. For all cases studied, the \emph{net} streamwise force is opposite to the vehicle's forward motion, i.e., it is a drag or breaking force. The streamwise force is, to first order, a linear function of the projected frontal area of the flapping amplitude $A^*=\sin{\beta}$, being generally insensitive to the flapping frequency. This linear relationship between the streamwise force and the amplitude $A^*$ is also observed for a quasi-steady fin. These two results support our hypothesis that the streamwise force produced by the fin can be interpreted as a pressure-based drag acting on a bluff body at high Reynolds number. For high flapping amplitudes and Strouhal numbers, a small instantaneous thrust production is observed during the final portion of the flapping cycle. Future studies may be able to reduce the time-average drag force produced by the flapping plate through the use of non-sinusoidal, asymmetric flapping kinematics, for example. Our data shows that the cross-stream forces scale with the amplitude $A^*$ and also with the Strouhal number $St$, unlike the streamwise forces that scale primarily with $A^*$ alone. We propose a piecewise scaling model for the cross-stream force that collapses the data very well over the entire range of parameters measured, indicating a regime change that is a function of the Strouhal number, with critical threshold around $St_c=0.08$ qualitatively changes the power scaling coefficient between below and above this threshold. We hypothesize that this deviation occurs due to the non-linear unsteady behavior of the flow over the fin at low angles. For sufficiently low fin angles during the upstroke phase of the motion,~\cite{he2023hydrodynamic} showed that the flow remains attached to the fin and to the lateral of the vehicle instead of producing vortices at the fin's trailing edge. This non-linearity greatly affects the quasi-steady cross-stream forces in the quasi-steady case, and likely also creates the two regimes observed for the lowest Strouhal compared to the higher Strouhal cases.

Finally, we demonstrate how the flapping fins can be used to maneuver an underwater vehicle of large inertial mass using a closed-loop control Cyber-physical system. Tests allowed the vehicle subject to an oncoming flow to maneuver in the lateral direction in response to the cross-stream forces produced by the flapping fins. The vehicle was constrained on all other degrees of freedom, providing an assessment of the cross-stream forces produced by the fins decoupled from other dynamic behaviors expected from a fully unconstrained vehicle or robot. The use of simultaneous flapping of the two fins located on opposite sides of the vehicle's main body allows for a fine control of its lateral position. 

In this study, the streamwise and cross-stream forces produced by a flapping fin were characterized for a stationary vehicle only. Future investigations addressing the impact of the lateral velocity on the force production should help the development of vehicle control strategies, predicting its dynamical behavior based on the velocity and acceleration of the vehicle, and the fin kinematics. Additionally, allowing the vehicle to move in the streamwise and yaw degrees of freedom will provide a complete picture of the capabilities of the flapping plates as bio-inspired control surfaces for underwater vehicles and robots.


\vspace{30pt}

\section*{Funding}
This research was funded by the Office of Naval Research, grant number N00014-21-1-2816.

\section*{Conflicts of Interest}
The authors declare no conflict of interest. 


\bibliographystyle{unsrt} 

\bibliography{refs}

\end{document}